\documentclass[fleqn,usenatbib]{mnras}

\usepackage{newtxtext,newtxmath}

\usepackage[T1]{fontenc}

\DeclareRobustCommand{\VAN}[3]{#2}
\let\VANthebibliography\thebibliography
\def\thebibliography{\DeclareRobustCommand{\VAN}[3]{##3}\VANthebibliography}

\usepackage{graphicx}	
\usepackage{amsmath}	

\title[Constraints on Short GRBs]{Constraints on Short Gamma-Ray Burst Physics and Their Host Galaxies from Systematic Radio Follow-up Campaigns}

\author[S. I. Chastain et al.]{S. I. Chastain,$^{1}$\thanks{E-mail: sarahchastain1@unm.edu (SIC)}
A. J. van der Horst,$^{2}$
G. E. Anderson,$^{3}$
L. Rhodes,$^{4}$
D. d'Antonio,$^{5,6,7}$
M. E. Bell,$^{5,8}$
\newauthor
R. P. Fender,$^{4,9}$ 
P. J. Hancock,$^{3,10}$
A. Horesh,$^{11}$
C. Kouveliotou,$^{2}$
K. P. Mooley,$^{12,13}$
A. Rowlinson,$^{14,15}$
\newauthor
S. D. Vergani,$^{16}$
R. A. M. J. Wijers,$^{13}$
and P. A. Woudt,$^{9}$
\\
$^{1}$Department of Physics and Astronomy, University of New Mexico, 210 Yale Blvd NE, Albuquerque, NM, 87106, USA\\
$^{2}$Department of Physics, George Washington University, 725 21st St NW, Washington, DC, 20052, USA\\
$^{3}$International Centre for Radio Astronomy Research, Curtin University, GPO Box U1987, Perth, WA 6845, Australia\\
$^{4}$Department of Physics, Astrophysics, University of Oxford, Denys Wilkinson Building, Keble Road, Oxford OX1 3RH, UK \\
$^{5}$University of Technology Sydney, 15 Broadway, Ultimo NSW 2007 \\
$^{6}$CSIRO, Space and Astronomy, PO Box 76, Epping, NSW 1710, Australia \\
$^{7}$Cognitivo, 11 York St, Sydney, NSW 2000, Australia \\
$^{8}$Leonardo.Ai Research Lab, 1 Kiara Cl, North Sydney, NSW 2060 \\
$^{9}$Department of Astronomy, University of Cape Town, Private Bag X3, Rondebosch 7701, South Africa\\
$^{10}$Curtin Institute for Data Science, Curtin University, GPO Box U1987, Perth, WA 6845, Australia\\
$^{11}$Racah Institute of Physics, The Hebrew University of Jerusalem, Jerusalem 91904, Israel\\
$^{12}$California Institute of Technology, 1200 E California Blvd, Pasadena, CA 91125\\
$^{13}$Indian Institute of Technology Kanpur, Kanpur 208016, U.P. India\\
$^{14}$Anton Pannekoek Institute for Astronomy, University of Amsterdam, Science Park 904, NL-1098 XH Amsterdam, Netherlands\\
$^{15}$ASTRON, the Netherlands Institute for Radio Astronomy, Postbus 2, NL-7990 AA Dwingeloo, Netherlands\\
$^{16}$DEPI, Observatoire de Paris, Universit\'e PSL, CNRS, 5 Place Jules Janssen, 92190, Meudon, France \\
}

\date{Accepted XXX. Received YYY; in original form ZZZ}

\pubyear{2024}

\begin{document}
\label{firstpage}
\pagerange{\pageref{firstpage}--\pageref{lastpage}}
\maketitle

\begin{abstract}
Short gamma-ray bursts (GRBs) are explosive transients caused by binary mergers of compact objects containing at least one neutron star. Multi-wavelength afterglow observations provide constraints on the physical parameters of the jet, its surrounding medium, and the microphysics of the enhanced magnetic fields and accelerated electrons in the blast wave at the front of the jet. The synchrotron radio emission can be tracked for much longer than in other spectral regimes, and it can pin down the evolution of the spectral peak. We present the results of a systematic observing campaign of eight short GRBs with the MeerKAT radio telescope. Additionally, we present observations of four of these short GRBs using the ATCA radio telescope and two of these short GRBs with the \textit{e}-MERLIN radio telescope. Using these results we report one possible detection of a short GRB afterglow from GRB 230217A and deep upper limits for the rest of our short GRB observations. We use these observations to place constraints on some of the physical parameters, in particular those related to electron acceleration, the circumburst density, and gamma-ray energy efficiency. We discuss how deeper observations with new and upgraded telescopes should be able to determine if the gamma-ray efficiency differs between long and short GRBs. We also report detections of the likely host galaxies for four of the eight GRBs and upper limits for another GRB, increasing the number of detected host galaxies in the radio with implications for the star formation rate in these galaxies.
\end{abstract}
\begin{keywords}
gamma-ray bursts -- radio continuum: transients -- radio continuum: galaxies
\end{keywords}



\section{Introduction}

Gamma-ray bursts (GRBs) are among the most powerful explosive transients in the Universe, accelerating electrons to extremely high Lorentz factors and emitting photons with up to TeV energies \citep[e.g.,][]{2019Natur.575..464A,2019Natur.575..459M}. These extreme sources can be divided into two varieties: long-soft and short-hard GRBs, based on the observed duration and spectral hardness of their prompt gamma-ray emission \citep{1993ApJ...413L.101K}. Long GRBs are caused by the collapse of massive stars \citep{1993ApJ...405..273W}, with nearby ones often associated with supernovae \citep[e.g.,][]{2003Natur.423..847H} \citep[see however,][]{2024Natur.626..737L,2022Natur.612..223R,2022Natur.612..228T}. Short GRBs are caused by binary mergers of compact objects \citep{1989Natur.340..126E}, with some of them associated with kilonovae \citep[e.g.,][]{2013Natur.500..547T}. In 2017, the electromagnetic emission of a short GRB was coincident with a gravitational wave event providing a wealth of information including conclusively linking binary neutron star mergers with short GRBs \citep{2017ApJ...848L..12A}. Following the GRB prompt gamma-ray emission, irrespective of the progenitor, an afterglow is observed across the electromagnetic spectrum, due to the interaction between the ejected material and the surrounding medium \citep{1998ApJ...497L..17S,1999ApJ...523..177W,1999ApJ...520L..29C}.

Multi-wavelength observations of GRBs provide constraints on the physical parameters of the collimated GRB outflow, i.e., the jet, its surrounding medium, and the microphysics of the enhanced magnetic fields and accelerated electrons in the blast wave at the front of the jet \citep{1998ApJ...497L..17S}. \citet{2014MNRAS.442.2342D}, for example, examined a sample of 36 short GRB X-ray observations from the 
{\it Neil Gehrels Swift Observatory} \citep[hereafter {\it Swift};][]{2004ApJ...611.1005G}
finding redshifts for 16 of the included GRBs and uncovering information about the binaries from which they originate. Radio observations provide a set of constraints that are particularly valuable such as tracking of the spectral breaks due to self-absorption and due to the minimum electron energy over time. Additionally, due to the power-law time evolution of the synchrotron emission and the ability to track the emission well into the non-relativistic phase, the radio emission of the afterglow can be tracked and well defined over a long period of time \citep[see, e.g.,][for a review]{2014PASA...31....8G}.

Radio detections of long GRBs are becoming increasingly common. \citet{chandra12} analysed all long GRBs that were observed in the radio band pre-dating 2011, resulting in a radio-detected fraction of $\sim30\%$. A more recent study that performed an unbiased radio follow-up campaign of 139 GRBs at a single frequency \citep[at 15.7 GHz using the Arcminute Microkelvin Imager Large Array;][]{zwart08} indicated that the detectable fraction could be higher at $\sim44-56$\%, and that radio afterglows can be detected within 24 hours post-burst \citep{anderson18}. 
The radio afterglows of short GRBs, on the other hand, are difficult to detect, with a little over a dozen detected \citep[] [note that this list does not include GRB 170817A/GW170817 as the jet was likely off-axis, with the radio afterglow rising at late times]{berger05,soderberg06,fong14,2015ApJ...815..102F,fong17gcn,2021ApJ...906..127F,lamb19,laskar22,2023arXiv230810936S,schroeder23GCN,2023GCN.35114....1S,2023GCN.35097....1R,2023GCN.35201....1R,anderson23GCN}.  
The radio afterglow of about half of this sample faded below detectability less than 2~days after the gamma-ray trigger \citep[e.g.][]{berger05,soderberg06,fong14,2021ApJ...906..127F,anderson23GCN,schroeder23GCN}; in these cases, the radio emission may be attributed to the reverse shock in the blast wave. 
In fact, all but 2 of the radio-bright short GRBs have been detected around 1 day after the trigger or earlier.
This motivates the need for rapid follow-up observations of short GRBs at radio frequencies \citep[e.g.][]{anderson21}. It is important to note that there is also a sample of short GRBs that remain detectable in the radio band for tens to hundreds of days \citep{lamb19,2021ApJ...906..127F,laskar22}, with one GRB even switching on at $\sim11$\,days, which was likely the result of late-time energy injection \citep{2023arXiv230810936S}. The latter suggests that previous radio monitoring campaigns of short GRBs that stop after one or two weeks may be missing late-time brightening or rebrightening episodes.

A larger and more complete sample of radio-monitored short GRBs needs to be compiled, covering multiple frequencies from very early (hours) to hundreds of days, in order to properly understand their environments, energetics, and jet properties. The most comprehensive study of the short GRB population thus far was conducted by \cite{2015ApJ...815..102F}, who used broadband modelling, including radio, to derive their physical parameters.
This study seems to suggest that the energetics of short GRBs is quite low, and that the same is true for the particle density in their environment, with densities as low as those found in the intergalactic medium, i.e., $10^{-6}$ cm$^{-3}$. \citet{2020MNRAS.495.4782O} discuss that such low densities may not be the correct interpretation of the observations and that, based on afterglow modelling, the host galaxy maybe misassociated resulting in an incorrect redshift determination. Furthermore, recent state-of-the-art modeling \citep{2022MNRAS.511.2848A} indicates that the energy of the blast wave is similar in long and short GRBs, which would mean that the difference in the observed gamma-ray energetics between short and long GRBs is due to differences in gamma-ray efficiency, while this efficiency is quite homogeneous in the population of long GRBs \citep{2016MNRAS.461...51B}. 

Various afterglow models \citep[e.g.,][]{2002ApJ...568..820G,2012ApJ...749...44V} have been used to derive physical GRB parameters by fitting multi-wavelength observations across various observing frequencies and timescales. Since the physical parameter space is large and complex, some methods have been derived to pin down one or two of the parameters using just a few observables. Most relevant for the work that we present here is that one can use the peak in radio light curves and spectra to constrain parameters related to electron acceleration in these sources \citep{2017MNRAS.472.3161B}. In this study, the authors determined typical values for the fraction $\epsilon_{\rm{e}}$ of the shock energy that goes into the population of accelerated electrons and established a fairly narrow distribution for this parameter. \citet{2023MNRAS.518.1522D} followed up on this work by additionally constraining the fraction $\xi_{\rm{e}}$ of electrons that gets accelerated by the blast wave into a power-law distribution of Lorentz factors, and the minimum Lorentz factor $\gamma_{\rm{m}}$ of this distribution of accelerated electrons. While this was done for a large sample of long GRBs, 49 to be exact, this has not been applied to a sample of short GRB radio afterglows.


These recent developments in modeling populations of GRBs, combined with a relatively small sample of short GRBs with radio detections \citep{2021ApJ...906..127F}, are a strong motivation to perform more radio studies of short GRBs. 
We have therefore carried out high-cadence radio monitoring of 8 short GRBs, combining observations taken with MeerKAT, the Australia Telescope Compact Array (ATCA), and \textit{e}-MERLIN. The MeerKAT observations were taken as part of the ThunderKAT project \citep{2016mks..confE..13F}, to systematically observe a sample of short GRBs. The ATCA observations form part of a program that utilizes the observatory's rapid-response mode to automatically trigger observations of short GRBs detected with \textit{Swift} \citep{anderson21}. This system enables ATCA to be on target within minutes of the GRB discovery (or when it has risen above the horizon) to target early radio afterglow emission seen to rise within 1~day post-burst in the radio-detected sample of short GRBs. 
While we do not significantly detect the radio counterpart of any short GRB in our sample, we are still able to put constraints on some physical properties of the GRB blast waves. In particular we place constraints on the gamma-ray efficiency, which is the efficiency of converting the total energy into gamma-ray emmision, and the electron density of the external medium. 

In this paper, we will use the MeerKAT, ATCA, and \textit{e}-MERLIN observations of a sample of short GRBs presented in section~\ref{sec:observations5} to find upper limits on possible GRB afterglows. In section~\ref{sec:methodandresult5} we present the results of our multi-epoch observations of 8 short GRB fields, and use the flux measurements and limits to constrain some GRB parameters and their environments. We also use our sensitive observations to constrain star formation rates in likely host galaxies of some of the GRBs in our sample. In section~\ref{sec:discussion5}, we discuss the implications of our measurements and place constraints on the gamma-ray efficiency and electron density of the external medium. We use these constraints to determine the possibility of future short GRB detections in radio with new large observatories. Finally, we provide a brief summary and conclusions in section~\ref{sec:conclusion5}.

\begin{table*}
\caption{Summary of our sample of MeerKAT short GRB observations at $\sim1.3$ GHz, their locations, position uncertainties, start time of each observation, $1\sigma$ rms noise in each image, and forced flux measurements at the GRB afterglow location (both integrated and peak flux).}
\label{tab:observations}
\begin{tabular}{lllllllll}
\hline
         GRB Name &        RA  &        Dec  & Pos. Unc.  & Position Ref.&  Days Post-Trigger &     RMS noise& $F_{\rm{int}}$  & $F_{\rm{peak}}$ \\
          &         (deg) &         (deg) & (arcsec) & &   &    ($\mu$Jy) &  ($\mu$Jy) &  ($\mu$Jy)\\
\hline
GRB200219A & $342.6384$ & $-59.1195$ & 1.8 & \citet{2020GCN.27138....1O} &      0.3$\;-\;$0.47 &  9 &  $31\pm13$ &      $29\pm7$ \\
GRB200219A & \textquotedbl & \textquotedbl & \textquotedbl & \textquotedbl &     2.21$\;-\;$2.38 &  7 &  $18\pm10$ &      $16\pm5$ \\
GRB200219A & \textquotedbl & \textquotedbl & \textquotedbl & \textquotedbl &      4.2$\;-\;$4.36 &  7 &  $31\pm10$ &      $30\pm6$ \\
GRB200219A & \textquotedbl & \textquotedbl & \textquotedbl & \textquotedbl &     8.27$\;-\;$8.44 &  8 &  $13\pm11$ &      $11\pm5$ \\
GRB200219A & \textquotedbl & \textquotedbl & \textquotedbl & \textquotedbl &  916.72$\;-\;$916.9 &  7 &  $19\pm10$ &      $17\pm5$ \\
GRB200411A &  $47.6642$ & $-52.3176$ & 1.5 & \citet{2020GCN.27538....1O} &     1.12$\;-\;$1.29 &  7 &  $24\pm11$ &      $22\pm6$ \\
GRB200411A &  \textquotedbl & \textquotedbl & \textquotedbl & \textquotedbl &      3.3$\;-\;$3.47 &  7 &  $52\pm11$ &      $51\pm8$ \\
GRB200411A &  \textquotedbl & \textquotedbl & \textquotedbl & \textquotedbl &      7.13$\;-\;$7.3 &  6 &  $37\pm10$ &      $36\pm6$ \\
GRB200411A &  \textquotedbl & \textquotedbl & \textquotedbl & \textquotedbl & 862.91$\;-\;$863.08 &  7 &  $35\pm10$ &      $34\pm6$ \\
GRB200522A &    $5.6818$ &  $-0.2827$ & 3.2 & \citet{2020GCN.27780....1B} &      0.81$\;-\;$0.99 & 22 &  $29\pm38$ &      $11\pm9$ \\
GRB200522A &   \textquotedbl & \textquotedbl & \textquotedbl & \textquotedbl &     1.77$\;-\;$1.95 & 27 &  $59\pm36$ &     $52\pm19$ \\
GRB200522A &   \textquotedbl & \textquotedbl & \textquotedbl & \textquotedbl &     6.61$\;-\;$6.79 & 20 &  $21\pm29$ &       $7\pm6$ \\
GRB200522A &   \textquotedbl & \textquotedbl & \textquotedbl & \textquotedbl &   14.61$\;-\;$14.78 & 29 &  $21\pm41$ &      $-6\pm7$ \\
GRB200522A &  \textquotedbl & \textquotedbl & \textquotedbl & \textquotedbl & 816.49$\;-\;$816.66 & 18 &  $23\pm26$ &      $13\pm8$ \\
GRB200907B &  $89.0290$ &   $6.9062$ & 1.8 & \citet{2020GCN.28391....1E} &     0.27$\;-\;$0.45 & 13 &   $9\pm23$ &     $-9\pm13$ \\
GRB200907B &  \textquotedbl & \textquotedbl & \textquotedbl & \textquotedbl &      2.3$\;-\;$2.48 & 12 &  $-9\pm26$ &     $17\pm30$ \\
GRB200907B &  \textquotedbl & \textquotedbl & \textquotedbl & \textquotedbl &     6.29$\;-\;$6.47 & 12 &  $-2\pm25$ &   $117\pm970$ \\
GRB200907B &  \textquotedbl & \textquotedbl & \textquotedbl & \textquotedbl &    17.32$\;-\;$17.5 & 12 &  $-4\pm26$ &    $54\pm208$ \\
GRB200907B &  \textquotedbl & \textquotedbl & \textquotedbl & \textquotedbl & 710.44$\;-\;$710.62 & 15 &  $-2\pm27$ &   $109\pm760$ \\
GRB210323A & $317.9472$ &  $25.3692$ & 1.6 & \citet{2021GCN.29703....1M} &     1.35$\;-\;$1.53 &  9 &  $26\pm14$ &      $23\pm7$ \\
GRB210323A & \textquotedbl & \textquotedbl & \textquotedbl & \textquotedbl &     3.34$\;-\;$3.52 &  9 &  $14\pm14$ &      $10\pm6$ \\
GRB210323A & \textquotedbl & \textquotedbl & \textquotedbl & \textquotedbl &     8.33$\;-\;$8.51 &  8 &  $15\pm13$ &      $11\pm6$ \\
GRB210323A & \textquotedbl & \textquotedbl & \textquotedbl & \textquotedbl &    11.72$\;-\;$11.9 &  9 &  $19\pm15$ &      $15\pm7$ \\
GRB210323A & \textquotedbl & \textquotedbl & \textquotedbl & \textquotedbl & 517.93$\;-\;$518.11 &  9 &  $15\pm14$ &      $11\pm6$ \\
GRB210919A &  $80.2546$ &   $1.3120$ & 0.5 & \citet{2021GCN.30883....1K} &     1.05$\;-\;$1.23 & 17 &  $-7\pm32$ &    $43\pm119$ \\
GRB210919A &  \textquotedbl & \textquotedbl & \textquotedbl & \textquotedbl &     5.11$\;-\;$5.29 & 18 &  $-9\pm30$ &     $24\pm45$ \\
GRB210919A &  \textquotedbl & \textquotedbl & \textquotedbl & \textquotedbl &     8.05$\;-\;$8.23 & 31 &  $-9\pm50$ &    $80\pm248$ \\
GRB210919A &  \textquotedbl & \textquotedbl & \textquotedbl & \textquotedbl &  335.2$\;-\;$335.38 & 18 & $-32\pm32$ &    $-21\pm12$ \\
GRB220730A & $225.0143$ & $-69.4959$ & 6.6 & \citet{2022GCN.32436....1D} &      2.12$\;-\;$2.3 &  8 &  $20\pm16$ &      $16\pm7$ \\
GRB220730A & \textquotedbl & \textquotedbl& \textquotedbl & \textquotedbl &     4.03$\;-\;$4.21 &  6 &   $0\pm10$ &    $31\pm177$ \\
GRB220730A & \textquotedbl & \textquotedbl& \textquotedbl & \textquotedbl &   10.98$\;-\;$11.16 &  6 &   $9\pm10$ &       $5\pm3$ \\
GRB220730A & \textquotedbl & \textquotedbl& \textquotedbl & \textquotedbl &   39.99$\;-\;$40.17 &  6 &    $3\pm9$ &     $-8\pm17$ \\
GRB220730A & \textquotedbl & \textquotedbl& \textquotedbl & \textquotedbl & 276.37$\;-\;$276.55 &  7 &  $15\pm11$ &      $13\pm5$ \\
GRB230217A & $280.7706$ & $-28.8379$ & 0.3 &      \citet{schroeder23GCN} &     5.17$\;-\;$5.34 &  7 &  $25\pm10$ &      $23\pm5$ \\
GRB230217A & \textquotedbl & \textquotedbl & \textquotedbl &      \textquotedbl &   10.16$\;-\;$10.33 &  8 &  $13\pm10$ &      $10\pm5$ \\
GRB230217A & \textquotedbl & \textquotedbl & \textquotedbl &      \textquotedbl &   20.26$\;-\;$20.43 &  7 &    $0\pm9$ & $-139\pm3590$ \\
GRB230217A & \textquotedbl & \textquotedbl & \textquotedbl &      \textquotedbl &   40.07$\;-\;$40.24 &  8 &    $1\pm9$ &    $-20\pm82$ \\
GRB230217A & \textquotedbl & \textquotedbl & \textquotedbl &      \textquotedbl & 155.78$\;-\;$155.95 &  8 &   $1\pm10$ &   $-24\pm105$ \\

\hline\end{tabular}
\end{table*}

\begin{table*}
\caption{Summary of our sample of ATCA short GRB observations, their locations, position uncertainties, the observing frequency, the start and end time of each observation, $1\sigma$ rms noise in each image, and forced flux measurements at the GRB afterglow location (both integrated and peak flux).}
\label{tab:ATCAobservations}
\begin{tabular}{llllllllll}
\hline
         GRB Name &        RA  &        Dec  & Pos. Unc.  & Position Ref.&  $\nu$ & Days Post-Trigger &     RMS noise& $F_{\rm{int}}$  & $F_{\rm{peak}}$ \\
          &         (deg) &         (deg) & (arcsec) & & (GHz)&   &    ($\mu$Jy) &  ($\mu$Jy) &  ($\mu$Jy)\\

\hline
GRB200219A & 342.6384 & -59.1196 & 1.8 & \citet{2020GCN.27138....1O} & 5.5 &   3.49$\;-\;$3.77 & 16 &  $15\pm36$ &    $-12\pm16$ \\
GRB200219A & \textquotedbl  & \textquotedbl & \textquotedbl & \textquotedbl & 9.0 &   3.49$\;-\;$3.77 & 15 &   $0\pm29$ & $587\pm20721$ \\
GRB200411A &  47.6642 & -52.3176 & 1.5 & \citet{2020GCN.27538....1O} & 5.5 &   0.07$\;-\;$0.28 & 12 &  $20\pm17$ &      $15\pm7$ \\
GRB200411A &  \textquotedbl & \textquotedbl & \textquotedbl & \textquotedbl & 9.0 &   0.07$\;-\;$0.28 & 10 &   $8\pm19$ &     $-7\pm10$ \\
GRB200411A &  \textquotedbl & \textquotedbl & \textquotedbl & \textquotedbl & 5.5 &   3.07$\;-\;$3.22 & 13 & $-14\pm22$ &      $-2\pm2$ \\
GRB200411A &  \textquotedbl & \textquotedbl & \textquotedbl & \textquotedbl & 9.0 &   3.07$\;-\;$3.22 & 13 &  $10\pm26$ &    $-14\pm21$ \\
GRB200411A &  \textquotedbl & \textquotedbl & \textquotedbl & \textquotedbl & 5.5 &   5.92$\;-\;$6.06 & 12 & $-13\pm21$ &      $-2\pm2$ \\
GRB200411A &  \textquotedbl & \textquotedbl & \textquotedbl & \textquotedbl & 9.0 &   5.92$\;-\;$6.06 & 11 &  $31\pm16$ &      $29\pm9$ \\
GRB200411A &  \textquotedbl & \textquotedbl & \textquotedbl & \textquotedbl & 5.5 & 11.07$\;-\;$11.19 & 16 & $-14\pm31$ &     $10\pm13$ \\
GRB200411A &  \textquotedbl & \textquotedbl & \textquotedbl & \textquotedbl & 9.0 & 11.07$\;-\;$11.19 & 15 & $-21\pm26$ &     $-10\pm7$ \\
GRB200907B &  89.0290 &   6.9062 & 1.8 & \citet{2020GCN.28391....1E} & 5.5 &  0.002$\;-\;$0.15  & 24 &       &   \\
GRB200907B &  \textquotedbl &   \textquotedbl & \textquotedbl & \textquotedbl & 9.0 &  0.002$\;-\;$0.15  & 18 &  &   \\
GRB200907B &  \textquotedbl &   \textquotedbl & \textquotedbl & \textquotedbl & 5.5 &   2.06$\;-\;$2.16 & 32 & &   \\
GRB200907B &  \textquotedbl &   \textquotedbl & \textquotedbl & \textquotedbl & 9.0 &   2.06$\;-\;$2.16 & 50 &   &  \\
GRB200907B &  \textquotedbl &   \textquotedbl & \textquotedbl & \textquotedbl & 5.5 &   6.18$\;-\;$6.31 & 40 &   &    \\
GRB200907B &  \textquotedbl &   \textquotedbl & \textquotedbl & \textquotedbl & 9.0 &   6.18$\;-\;$6.31 & 28 &    &       \\
GRB200907B &  \textquotedbl &   \textquotedbl & \textquotedbl & \textquotedbl & 5.5 &  12.06$\;-\;$12.12 & 30 &    &      \\
GRB200907B &  \textquotedbl &   \textquotedbl & \textquotedbl & \textquotedbl & 9.0 & 12.06$\;-\;$12.12 & 40 &   &   \\
GRB220730A & 225.0143 & -69.4959 & 6.6 & \citet{2022GCN.32436....1D} & 5.5 &   0.38$\;-\;$0.87 & 12 & $-12\pm28$ &     $11\pm15$ \\
GRB220730A & \textquotedbl & \textquotedbl & \textquotedbl & \textquotedbl & 9.0 &   0.38$\;-\;$0.87 &  7 &  $18\pm19$ &      $12\pm7$ \\
GRB220730A & \textquotedbl & \textquotedbl & \textquotedbl & \textquotedbl & 5.5 &   1.72$\;-\;$1.86 & 20 & $-47\pm49$ &    $-30\pm18$ \\
GRB220730A & \textquotedbl & \textquotedbl & \textquotedbl & \textquotedbl & 9.0 &   1.72$\;-\;$1.86 & 11 &  $25\pm25$ &     $17\pm10$ \\
GRB220730A & \textquotedbl & \textquotedbl & \textquotedbl & \textquotedbl & 5.5 &   3.58$\;-\;$3.84 & 16 & $-16\pm41$ &     $18\pm26$ \\
GRB220730A & \textquotedbl & \textquotedbl & \textquotedbl & \textquotedbl & 9.0 &   3.58$\;-\;$3.84 & 10 &  $36\pm26$ &     $29\pm12$ \\
GRB220730A & \textquotedbl & \textquotedbl & \textquotedbl & \textquotedbl & 5.5 & 10.43$\;-\;$10.72 & 15 & $-16\pm33$ &       $6\pm7$ \\
GRB220730A & \textquotedbl & \textquotedbl & \textquotedbl & \textquotedbl & 9.0 & 10.43$\;-\;$10.72 &  9 &  $12\pm23$ &      $-4\pm5$ \\
\hline
\end{tabular}
\end{table*}
\begin{table*}
\caption{Summary of our sample of \textit{e}-MERLIN short GRB observations, their locations, position uncertainties, the observing frequency, the start and end time of each observation, and $1\sigma$ rms noise in each image. 
}
\label{tab:eMerlinobservations}
\begin{tabular}{llllllll}
\hline
         GRB Name &        RA  &        Dec  & Pos. Unc.  & Position Ref.&  $\nu$ & Days Post-Trigger &     RMS noise\\
          &         (deg) &         (deg) & (arcsec) & & (GHz)&   &    ($\mu$Jy) \\

\hline

GRB200907B &  89.0290 &   6.9062 & 1.8 & \citet{2020GCN.28391....1E} & 4.8 &  1.27$\;-\;$1.93 &28 \\
GRB210323A & 317.9472 &  25.3692 & 1.6 & \citet{2021GCN.29703....1M} & 4.8  & 1.27$\;-\;$1.63 &17\\
GRB210323A & \textquotedbl & \textquotedbl & \textquotedbl & \textquotedbl & 4.8  & 2.16$\;-\;$2.75 &17\\
GRB210323A & \textquotedbl & \textquotedbl & \textquotedbl & \textquotedbl & 4.8  & 3.16$\;-\;$3.75 &17\\
GRB210323A & \textquotedbl & \textquotedbl & \textquotedbl & \textquotedbl & 4.8  & 4.16$\;-\;$4.75 &15\\
GRB210323A & \textquotedbl & \textquotedbl & \textquotedbl & \textquotedbl & 4.8  & 5.60$\;-\;$5.90 &20\\
GRB210323A & \textquotedbl & \textquotedbl & \textquotedbl & \textquotedbl & 4.8  & 8.15$\;-\;$8.50 &24\\
GRB210323A & \textquotedbl & \textquotedbl & \textquotedbl & \textquotedbl & 4.8  & 9.21$\;-\;$9.32 &34\\
GRB210323A & \textquotedbl & \textquotedbl & \textquotedbl & \textquotedbl & 4.8  & 10.21$\;-\;$10.32&27\\

\hline
\end{tabular}
\end{table*}
\section{Observations and Data Analysis}
~\label{sec:observations5}
\subsection{MeerKAT Observations}\label{sec:observations_meerkat}
The MeerKAT radio telescope \citep[MeerKAT;][]{2016mks..confE...1J} is a radio interferometer in South Africa that will be integrated into the middle frequency component of the Square Kilometer Array~\citep[SKA;][]{2009IEEEP..97.1482D}.
The aim of the project presented here was to observe short GRBs, all of which were detected with {\it Swift}, that had follow-up observations in at least one other waveband (X-rays, ultraviolet, optical and/or near-infrared) and are visible to the MeerKAT radio telescope. The observing campaign started in early 2020, with the last observations in July 2023. A total of 9 GRBs met this criteria. Of these, GRB~210726A is presented in~\citet{2023arXiv230810936S} and not included in this work, thus leaving a total of 8 GRBs presented here. We performed multi-epoch observations of these 8 GRBs, spanning days to weeks, and in 6 cases even years (the 7th GRB occurred in July 2022 and the 8th in May 2023). This resulted in 37 total observations for our sample of 8 short GRBs, with a central frequency of 1.3~GHz. All observations lasted approximately 4 hours, resulting in $1\sigma$ noise levels down to $7-9\,\mu$Jy in most fields. In some cases the image noise was significantly higher, up to a few tens of $\mu$Jy, due to bright sources causing strong artifacts. The observing details and results are summarized in Table~\ref{tab:observations}. 

Each observation was processed using the \textsc{ProcessMeerKAT} pipeline \citep{pminprep} for calibration. The imaging was performed with Common Astronomy Software Applications \citep[\textsc{CASA};][]{2022arXiv221002276T} using the task \textit{tclean}. An initial shallow image was made, then self-calibration and refined flagging for radio-frequency interference was performed, followed by making a final, deep image. W-projection with 128 w-projection planes was used to correct for the non-coplanar baselines. The multi-term multi-scale imaging algorithm was used with 2 Taylor terms and at least three different scales, always including 0, 5, and 15. 

Even though the vast majority of our observations resulted in deep images, none of the ones presented here had significant detections at the GRB location. Some of the observations resulted in possible detections of a host galaxy, which will be discussed in Section~\ref{sec:methodandresult5}.  The short GRB~210726A was also observed as part of our campaign, resulting in some detections, which are presented as part of a multi-wavelength study in \citet{2023arXiv230810936S}. 

\begin{figure*}
    \includegraphics[width=\textwidth]{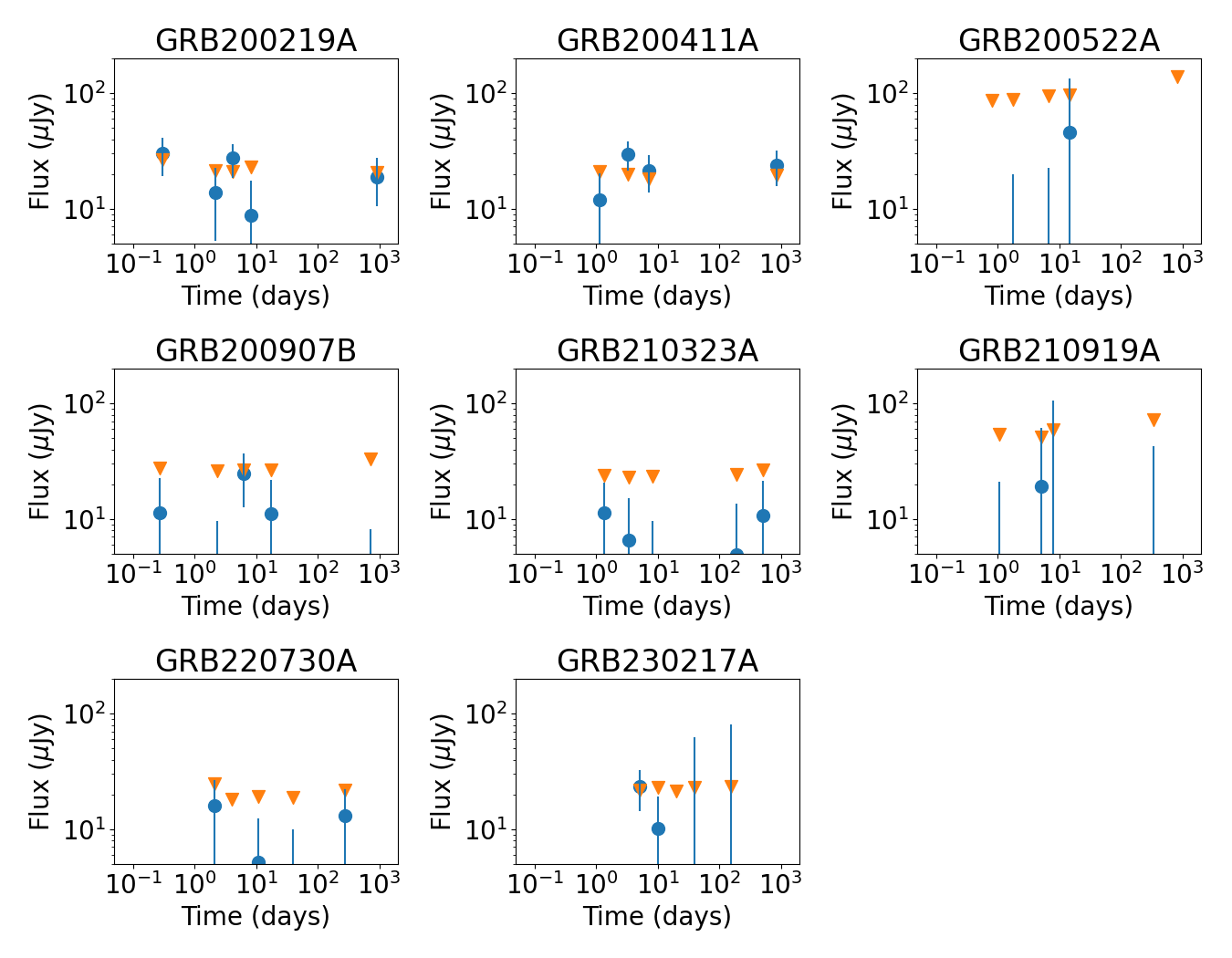}
    \caption{ MeerKAT observations of the 8 short GRBs in our sample, with forced peak flux measurements shown as blue circles and $3\sigma$ upper limits based on the image rms noise shown as orange triangles.}
    \label{fig:allmeasurements}
\end{figure*}

Given the non-detections of the GRBs in our sample, forced flux measurements at the GRB location were performed with the LOFAR transients pipeline \citep[TraP;][]{2015A&C....11...25S} by supplying a fixed location of the short GRB afterglow along with using standard parameter settings; the image noise was also measured with the TraP. The light curves of the forced peak flux measurements along with $3\sigma$ upper limits are shown in Figure~\ref{fig:allmeasurements}. 

\begin{figure*}
    \includegraphics[width=\textwidth]{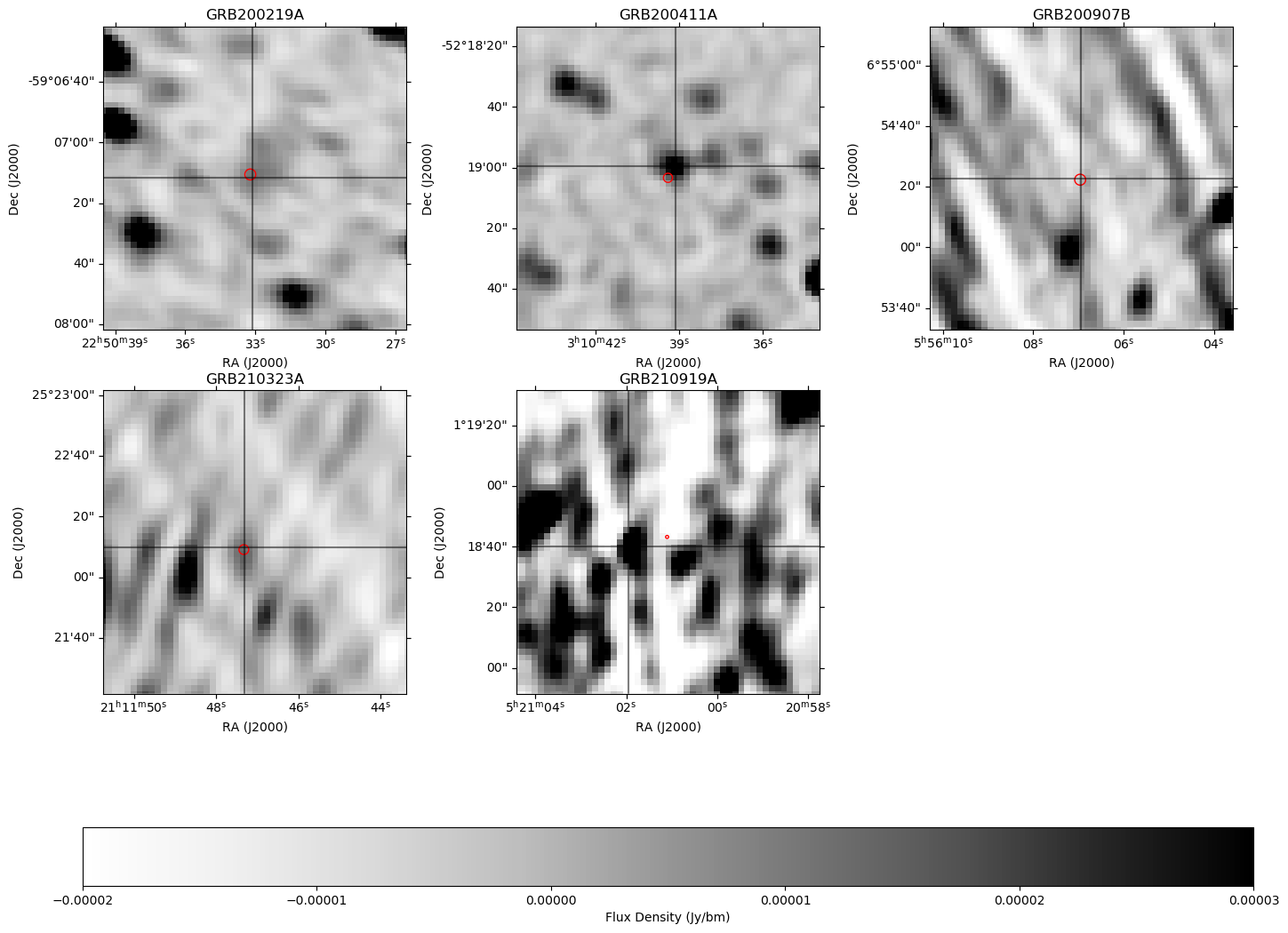}
    \caption{A combined image of all of our MeerKAT observations for the GRBs in which there is a host galaxy detection reported. The GRB localization is indicated by a red circle and the host galaxy location is marked with the crosshair. Note how close and bright the host galaxy is in comparison to the GRB location for GRB 200411A.}
    \label{fig:GRB200411Afield}
\end{figure*}

Note that almost all the forced peak flux measurements for GRB~200411A are higher than the $3\sigma$ upper limits because the host galaxy, which is offset from the GRB position by only 4.6 arcseconds and constant in time, is contributing to the flux at that position due to the restoring beam size being between 6 to 9 arcseconds. Figure~\ref{fig:GRB200411Afield} shows combined images of the observed GRBs with host galaxy counterparts, excluding GRB~200522A due to artefacts from bright sources in the field. The crosshair marks the host galaxy location and the red circle marks the localization of the GRB afterglow. The host galaxy of GRB~200411A is close to the GRB localization and appears to be contributing to the flux measured at the GRB location. Therefore, the flux measured from this GRB is from this galaxy, which is likely its host galaxy, and not from the GRB afterglow. We investigate these host galaxies further in section~\ref{sec:methodandresult5}.

\subsection{ATCA Observations}

ATCA is a 6-dish interferometer with a maximum baseline length of 6\,km based in New South Wales, Australia. 
Since 2018, we have been using ATCA to trigger rapid-response and monitoring observations of short GRBs under project code C3204 \citep[PI Anderson;][]{anderson21}. For all observations, we use the dual 4cm receiver with central frequencies of 5.5 and 9\,GHz, each having a 2\,GHz bandwidth.  
The program followed up four of the eight short GRBs monitored by MeerKAT, with the details listed in Table~\ref{tab:ATCAobservations}.  
Any triggered observations take place within 1\,day post-burst and are then followed by up to three quasi-logarithmically spaced, manually scheduled observations over the next $\sim2-3$ weeks.
The first observations of GRB 200411A and GRB 200907B were taken using the rapid-response observing mode, with ATCA being on target and observing GRB 200907B just 3 minutes post-burst. 

\begin{figure*}
\includegraphics[width=\textwidth]{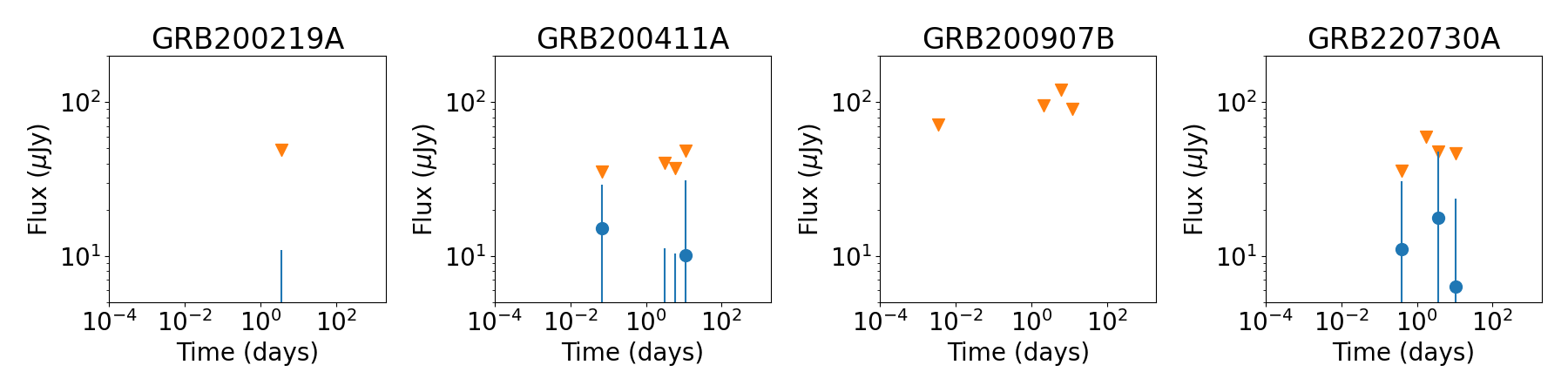}
    \includegraphics[width=\textwidth]{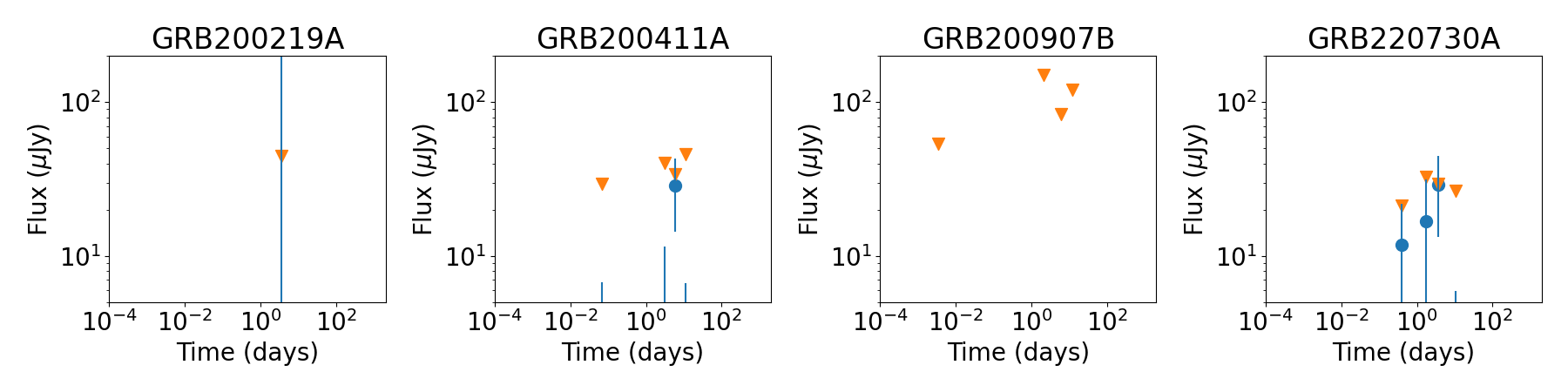}
    \caption{ATCA observations at 5.5 GHz (top) and 9 GHz (bottom) of four of the short GRBs in our sample, with forced peak flux measurements shown as blue circles and $3\sigma$ upper limits based on the image rms noise shown as orange triangles.}
    \label{fig:allATCA9measurements}
\end{figure*}

The observations were calibrated in \textsc{CASA} by first using the task importatca to convert the RPFITS data format into measurement sets. The data were then calibrated using standard \textsc{CASA} tasks to flag, generate calibration tables, and apply calibration tables. After calibrating the data, the \textsc{CASA} task \textit{tclean} was then used to make images of the field. In all cases, the standard gridder and default Hogbom deconvolution algorithm were used. For the GRB 200219A and GRB 220730A fields, multiple rounds of self-calibration were necessary due to a bright source in the field. The initial round of cleaning used a relatively shallow threshold followed by phase-only calibration. The following rounds were cleaned progressively deeper and after a few rounds of phase-only calibration, phase and amplitude calibration was performed. The GRB 200411A field did not require self-calibration. 
The GRB 200907B field was calibrated and imaged using Miriad~\citep{1995ASPC...77..433S} using standard techniques. However, we were unable to perform a forced-fit to the GRB location in TraP due to the synthesised beam being too elongated. Therefore we report only the rms noise.
The force-fitted flux density measurements listed in Table~\ref{tab:ATCAobservations} were performed in the same way as was described in Section~\ref{sec:observations_meerkat}, with the 5.5\,GHz and 9\,GHz measurements plotted in Figure~\ref{fig:allATCA9measurements}. Note that although we appear to have detected the GRB afterglow in the third observation of GRB 220730A, this is likely due to a sidelobe from a nearby bright source.

    

\begin{figure}
\includegraphics[width=\columnwidth]{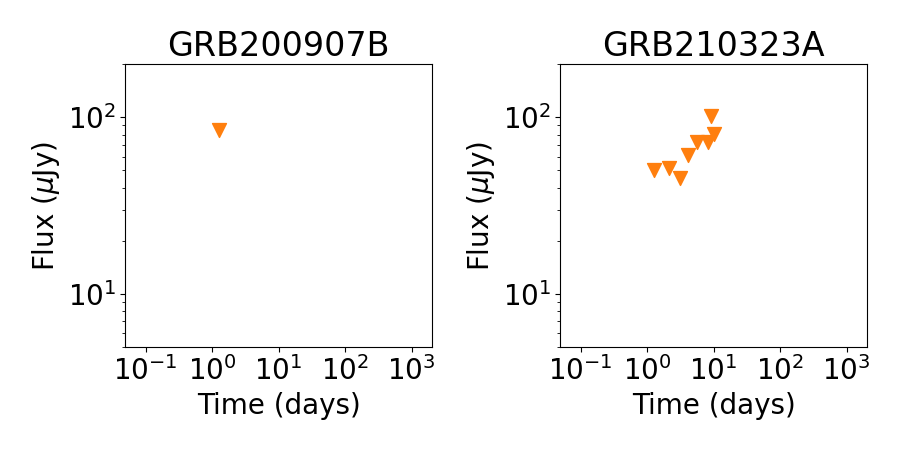}
    \caption{\textit{e}-MERLIN observations at 4.8 GHz of two of the short GRBs in our sample, with $3\sigma$ upper limits based on the image rms noise shown as orange triangles.}
    \label{fig:allemerlinmeasurements}
\end{figure}
\subsection{\textit{e}-MERLIN Observations}

The \textit{enhanced} Multi-Element Remotely Linked Interferometer Network (\textit{e}-MERLIN) is a UK-based radio facility consisting of seven antennas with a longest baseline of 217~km. Our observations were taken under the long running program ``High-resolution observations of short GRBs beyond the LIGO horizon'' (CY10002, PI: Rhodes). We observed two short GRBs: GRB 200907A and 210323A. GRB 200907A was observed for one epoch on 8\textsuperscript{th} September 2020. GRB 210323A was observed for eight epochs between 25 March and 3 April 2021. The observations were made at 4.8\,GHz. Each observation was reduced using the \textit{e}-MERLIN pipeline \citep{moldon2021}. The pipeline performs flagging for radio frequency interference and applies apriori flags including antenna shadowing. It then averages down the data for calibration. Bandpass calibration is performed using 1407$+$2827 (Mrk 668) as the calibrator followed by complex gain calibration with 1331$+$3030 (3C286) and J2114$+$2832 as the flux density and phase calibrators, respectively. 

On receiving the data, we use \textsc{CASA} (Version 5.8.0) to perform further flagging and image the data. A full summary of the observations is given in Table~\ref{tab:eMerlinobservations} with the limits plotted in Figure~\ref{fig:allemerlinmeasurements}. For these observations we only report the rms noise which was measured by TraP due to the resolution of the instrument being much finer than the localization region of the GRB afterglow.

\section{Modeling Results and Host Galaxies}
~\label{sec:methodandresult5}

In this section we put our results in the context of GRB modeling and host galaxy studies.

\subsection{Constraining GRB Afterglow Physics}

We use our deep MeerKAT, ATCA and \textit{e}-MERLIN observations to constrain some of the physical parameters of GRB afterglows for our sample of short GRBs. One approach could be to perform multi-wavelength modeling covering a wide range of timescales and frequencies, but the amount of data available for this is limited. The reason for this is two-fold: (1) short GRBs are typically faint across the electromagnetic spectrum, and that is also true for the GRBs in our sample; and (2) due to the global COVID-19 pandemic, there were limited observations, in particular in the optical, during a significant fraction of our follow-up campaign. Therefore, we focus here on the MeerKAT, ATCA and \textit{e}-MERLIN observations, and what can be learned from the deep limits we obtained on their radio brightness.

We model our MeerKAT observations by comparing the $3\sigma$ upper limits to the theoretical expectations of peaks in radio light curves and spectral energy distributions from \citet{2017MNRAS.472.3161B} and \citet{2023MNRAS.518.1522D}. For every observation, we can consider the measured upper limit to be a limit on the theoretical peak flux if it would be at that time and frequency. Given the quasi-logarithmic spacing of our observations, similar flux limits at various times for a given GRB, and the power-law flux evolution of GRB afterglows, it is unlikely that we would have missed the radio peak in the gaps between our observations, except possibly for the observations between a few weeks and one or two years after the GRB as can be seen by a slight date gap in measurements in Figure~\ref{fig:allmeasurements}. 

For every observation, we calculate the $\Psi$ parameter from \citet{2017MNRAS.472.3161B} and  \citet{2023MNRAS.518.1522D}. The $\Psi$ parameters is a proxy for $\epsilon_e$ and $\xi_{\rm{e}}$, and can be expressed as a combination of observables related to the peak in the radio light curve. For reference we reproduce the $\Psi$ equation for the homogeneous environment (which is expected for short GRBs):
\begin{align}
    \Psi &= \left(\frac{ 261.4 (1+z)^{1/2} \nu_p t_p^{3/2} E_{\gamma,iso,53}^{1/2}}{10^{15} d_{28}^2 F_{\nu_p} \text{max}(1,t_p/t_j)^{1/2}}\right)^{1/2} \\
    &= \frac{(p-2)}{0.177(p-1)} \left(\frac{p-0.67}{p+0.14}\right)^{1/2} \left(\frac{1-\epsilon_{\gamma}}{\epsilon_{\gamma}}\right)^{-1/4} n_0^{-1/4} \epsilon_e \xi_e^{-3/2}
\end{align}
This parameter is derived from the ratio of the theoretical equations for the peak flux and peak frequency. It can be expressed in terms of observables and in terms of physical parameters, thus allowing us to draw conclusions about the physical parameters using our observations. The observables are the peak observing frequency, $\nu_p$, peak observation time $t_p$, the jet break time, $t_j$, peak flux, $F_{\nu_p}$, redshift, $z$, luminosity distance, $d_{28}$, and isotropic-equivalent gamma-ray energy, $E_{\gamma,iso,53}$. We show the redshift and isotropic equivalent gamma-ray energy values used, and their references, in Table~\ref{tab:GRBprops}. For the 2 GRBs without a known redshift, we use the average redshift of the other 6 GRBs. The luminosity distance of each GRB was calculated using \citet{2006PASP..118.1711W}, and we assume that the peak time, $t_p$, occurs before the jet break time, $t_j$. 

The parameter $\Psi$ strongly depends on two parameters related to the electron acceleration process: the fraction $\epsilon_{\rm{e}}$ of the shock energy that goes into the population of accelerated electrons, and the fraction $\xi_{\rm{e}}$ of electrons that gets accelerated by the blast wave into a power-law distribution of Lorentz factors: $\Psi\propto\epsilon_{\rm{e}}\;\xi_{\rm{e}}^{-3/2}$. It weakly depends on the power-law slope $p$ of the Lorentz factor distribution, the density $n_0$ of the circumburst medium, and the gamma-ray efficiency, $\epsilon_{\gamma}$. We note that we are using the $\Psi$ equations from \citet{2023MNRAS.518.1522D} for a homogeneous medium, since that is the typically assumed environment of a short GRB (rather than a stellar wind, which one may expect for long GRBs).

\begin{figure}
	\includegraphics[width=\columnwidth]{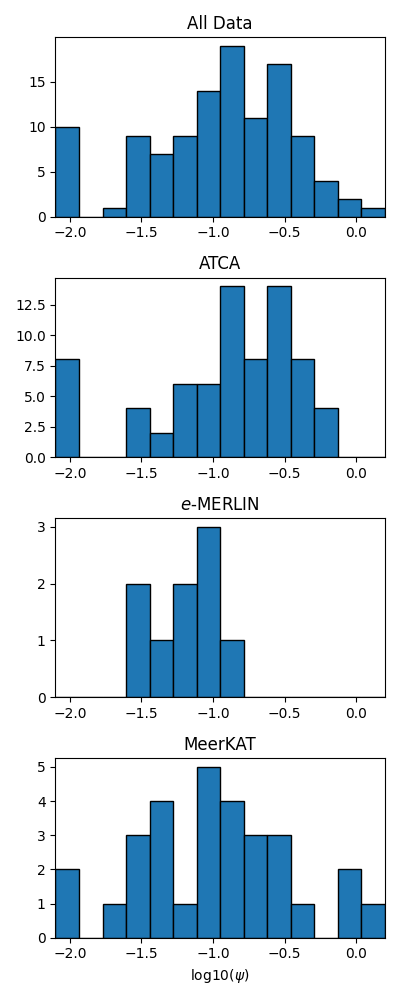}
    \caption{Histogram of lower limits on $\Psi$ for all measurements within the first 40 days after the gamma-ray trigger.}
    \label{fig:psiobshist}
\end{figure}

We use the $3\sigma$ upper limits on the peak flux to estimate lower limits on the $\Psi$ parameter, which are shown in Tables~\ref{tab:psiobs1}, \ref{tab:psiobs2}, and \ref{tab:psiobs3}. In this analysis we treat GRB 230217A as upper limits. The possible detection of GRB 230217A is at approximately $3\sigma$ and occurs in the first observation. The limits for the $\Psi$ parameter are calculated only for observations in the first 40 days since this methodology only holds for a relativistic GRB blast wave. In Figure~\ref{fig:psiobshist} we show histograms of the lower limits on $\Psi$ using all of our observations, the MeerKAT observations only, the \textit{e}-MERLIN observations only, and the ATCA observations only, and in Table~\ref{tab:averagepsi} we report the average upper limits in log-space on $\Psi$ for each of these sets of observations along with values from \citet{2017MNRAS.472.3161B} and \citet{2023MNRAS.518.1522D}. This distribution of lower limits on $\Psi$ can be compared to the distribution of estimated $\Psi$ values for a large sample of long GRBs \citep{2023MNRAS.518.1522D}; the implications of this are discussed in section~\ref{sec:discussion5}.

\begin{table}
    \caption{The log-space mean values of the upper limits on $\Psi$ for the observations from each telescope in this study, for all telescopes combined in this study, for \citet{2017MNRAS.465.4106C} ISM and wind environments, and for \citet{2023MNRAS.518.1522D} ISM and wind environments. Note that the \citet{2023MNRAS.518.1522D} are weighted by uncertainties.}
    \label{tab:averagepsi}
    \begin{tabular}{ll}
    \hline
    Dataset & Average $\Psi$ upper limit  \\
    \hline
    Combined & 0.1 \\ 
    MeerKAT & 0.1 \\
    ATCA & 0.13\\
    \textit{e}-MERLIN & 0.07 \\
    \citet{2017MNRAS.472.3161B} ISM & 0.15 \\
    \citet{2017MNRAS.472.3161B} Wind & 0.13 \\
    \citet{2023MNRAS.518.1522D} ISM & 0.11\\
    \citet{2023MNRAS.518.1522D} Wind & 0.14\\
    \end{tabular}
\end{table}

\begin{table*}
\caption{Trigger date, redshift, isotropic-equivalent gamma-ray energy for the short GRBs in our sample, including references. For GRBs without a known redshift, the average of the other redshifts is used: 0.533.}
\label{tab:GRBprops}
\begin{tabular}{l|lll|ll}
\hline
        GRB & Trigger Date & Redshift &  &        $E_{\gamma,iso}$ &  \\
        \hline
        GRB200219A & 2020-02-19 7:36:49 &$0.48\pm0.02$ &  \citet{2022ApJ...940...56F} \ &$3.74\times10^{51}$ & \citet{2020GCN.27226....1S}\\
        GRB200411A & 2020-04-11 4:29:02 &$0.6\pm0.1$ & \citet{2022MNRAS.515.4890O} &$7.17\times10^{50}$ & \citet{2020GCN.27543....1B}\\
        GRB200522A & 2020-05-22 11:41:34 &$0.5536\pm0.0003$ & \citet{2016arXiv161205560C} &$1.39\times10^{50}$ & \citet{2020GCN.27793....1U} \\
        \textquotedbl & \textquotedbl & \textquotedbl & \citet{2021ApJ...906..127F} & \textquotedbl & \textquotedbl  \\
        GRB200907B & 2020-09-07 18:51:11 &$0.56^{+1.39}_{-0.32}$ & \citet{2022ApJ...940...56F} &$1.28\times10^{50}$ & \citet{2020GCN.28398....1K}\\
        GRB210323A & 2021-03-23 22:02:18&$0.733\pm0.001$ & \citet{2022ApJ...940...56F}  &$2.18\times10^{50}$ & \citet{2021GCN.29713....1F} \\
        GRB210919A & 2021-09-19 00:28:34 &$0.27\pm0.12$ & \citet{2021GCN.30934....1O} &$5.53\times10^{49}$ & \citet{2022GCN.31566....1M} \\
        GRB220730A & 2022-07-30 15:48:55 & &  &$9.09\times10^{50}$ & \citet{2022GCN.32439....1F} \\     
        GRB230217A &2023-02-17 21:53:11 & &  &$1.55\times10^{52}$ & \citet{2023GCN.33353....1V} \\     

\end{tabular}
\end{table*}

\begin{table}
\caption{Lower limits on $\Psi$ estimated from $3\sigma$ upper limits on the radio flux in MeerKAT observations at $\sim1.3$~GHz. In this analysis we treat GRB 230217A observations as upper limits. The detection of GRB 230217A is at approximately $3\sigma$ and occurs in the first observation. Note that the $\Psi$ parameter is calculated only for observations in the first 40 days since this methodology only holds for a relativistic GRB blast wave.}
\label{tab:psiobs1}
\begin{tabular}{llll}
\hline
        GRB & Days       & Flux upper  & $\psi$ \\
            & post-burst & limit ($\mu$Jy) & lower limit \\
        \hline

GRB200219A &   0.3 & 27 & 0.03 \\
GRB200219A &  2.21 & 21 & 0.13 \\
GRB200219A &   4.2 & 21 & 0.21 \\
GRB200219A &  8.27 & 23 & 0.34 \\
GRB200411A &  1.12 & 21 & 0.04 \\
GRB200411A &   3.3 & 20 & 0.09 \\
GRB200411A &  7.13 & 18 & 0.18 \\
GRB200522A &  0.81 & 87 & 0.01 \\
GRB200522A &  1.77 & 89 & 0.02 \\
GRB200522A &  6.61 & 96 & 0.05 \\
GRB200522A & 14.61 & 97 & 0.09 \\
GRB200907B &  0.27 & 28 & 0.01 \\
GRB200907B &   2.3 & 26 & 0.04 \\
GRB200907B &  6.29 & 27 & 0.09 \\
GRB200907B & 17.32 & 27 &  0.2 \\
GRB210323A &  1.35 & 24 & 0.03 \\
GRB210323A &  3.34 & 23 & 0.05 \\
GRB210323A &  8.33 & 24 &  0.1 \\
GRB210323A & 11.72 & 24 & 0.13 \\
GRB210919A &  1.05 & 54 & 0.03 \\
GRB210919A &  5.11 & 51 &  0.1 \\
GRB210919A &  8.05 & 60 & 0.14 \\
GRB220730A &  2.12 & 25 & 0.07 \\
GRB220730A &  4.03 & 18 & 0.14 \\
GRB220730A & 10.98 & 19 & 0.29 \\
GRB220730A & 39.99 & 19 & 0.76 \\
GRB230217A &  5.17 & 22 & 0.31 \\
GRB230217A & 10.16 & 23 &  0.5 \\
GRB230217A & 20.26 & 21 & 0.87 \\
GRB230217A & 40.07 & 23 & 1.41 \\
\end{tabular}
\end{table}

\begin{table}
\caption{Lower limits on $\Psi$ estimated from $3\sigma$ upper limits on the radio flux in ATCA observations.}
\label{tab:psiobs2}
\begin{tabular}{lllll}
\hline
      GRB & Days& $\nu_{p}$        & Flux upper  & $\psi$ \\
            & post-burst & (GHz) & limit ($\mu$Jy) & lower limit \\
        \hline
GRB200219A &  3.49 &  5.5 & 49 & 0.25 \\
GRB200219A &  3.49 &  9.0 & 45 & 0.34 \\
GRB200411A &  0.07 &  5.5 & 35 & 0.01 \\
GRB200411A &  3.07 &  5.5 & 40 & 0.13 \\
GRB200411A &  5.92 &  5.5 & 37 & 0.22 \\
GRB200411A & 11.07 &  5.5 & 49 & 0.31 \\
GRB200411A &  0.07 &  9.0 & 29 & 0.01 \\
GRB200411A &  3.07 &  9.0 & 40 & 0.17 \\
GRB200411A &  5.92 &  9.0 & 34 &  0.3 \\
GRB200411A & 11.07 &  9.0 & 46 & 0.41 \\
GRB200907B &  2.06 &  5.5 & 79 & 0.05 \\
GRB200907B &  6.18 &  5.5 & 96 &  0.1 \\
GRB200907B &  2.06 &  9.0 & 58 & 0.07 \\
GRB200907B &  6.18 &  9.0 & 78 & 0.14 \\
GRB220730A &  0.38 &  5.5 & 36 & 0.03 \\
GRB220730A &  1.72 &  5.5 & 60 & 0.08 \\
GRB220730A &  3.58 &  5.5 & 48 & 0.16 \\
GRB220730A & 10.43 &  5.5 & 46 & 0.37 \\
GRB220730A &  0.38 &  9.0 & 21 & 0.06 \\
GRB220730A &  1.72 &  9.0 & 33 & 0.15 \\
GRB220730A &  3.58 &  9.0 & 29 & 0.27 \\
GRB220730A & 10.43 &  9.0 & 26 & 0.62 
\end{tabular}
\end{table}

\begin{table}
\caption{Lower limits on $\Psi$ estimated from $3\sigma$ upper limits on the radio flux in \textit{e}-MERLIN observations.}
\label{tab:psiobs3}
\begin{tabular}{lllll}
\hline
      GRB & Days& $\nu_{p}$        & Flux upper  & $\psi$ \\
            & post-burst & (GHz) & limit ($\mu$Jy) & lower limit \\
        \hline
GRB200907B &  1.27 &  4.8 &  85 & 0.03 \\
GRB210323A &  1.27 &  4.8 &  50 & 0.03 \\
GRB210323A &  2.17 &  4.8 &  51 & 0.05 \\
GRB210323A &  3.17 &  4.8 &  45 & 0.07 \\
GRB210323A &  4.17 &  4.8 &  61 & 0.07 \\
GRB210323A &   5.6 &  4.8 &  72 & 0.08 \\
GRB210323A &  8.15 &  4.8 &  72 & 0.11 \\
GRB210323A &  9.21 &  4.8 & 103 &  0.1 \\
GRB210323A & 10.21 &  4.8 &  81 & 0.13 \\
\end{tabular}
\end{table}

\subsection{Constraining Host Galaxy Star Formation Rates}

While we do not have significant detections of the GRBs presented here, our images include many other sources, including some steady sources very nearby GRB locations. Given that many short GRBs reside in the outskirts or possibly even outside of their host galaxy, we explored the possibility that some of these nearby sources are associated with the potential host galaxies. For 6 GRBs in our sample, host galaxy locations in optical images have been published. One of these GRBs is GRB 200522A, which we omit from the host galaxy analysis, since the MeerKAT image has significant artefacts from two very bright sources in the field of view. For all of the other sources we give the MeerKAT location and peak flux in Table~\ref{tab:GRBhostSFR}. Since we expect the host galaxies to be relatively constant in flux, and in order to improve sensitivity and \textit{uv} coverage, the fluxes were measured in deep combined images of each field (12 to 16 hours of observing time in total for each field), by force fitting the flux at the host galaxy location using the TraP. We show these combined images along with the GRB localization, marked as a red circle, and host galaxy location, marked as a crosshair in Figure~\ref{fig:GRB200411Afield}. In four of the five cases, we have a $\gtrapprox3\sigma$ detection, with the exception being GRB~200907B, in which case we adopt a $3\sigma$ upper limit. In all four of these cases, the source at the location of the host galaxy is either significantly offset from the afterglow position or is not variable even on timescales beyond a year. With these detections, the number of detected short GRB host galaxies is significantly increased. \citet{2019ApJ...887..206K} carried out a dedicated search at 5.5 GHz and 9.0 GHz with the Karl G. Jansky Very Large Array (VLA) \citep{2011ApJ...739L...1P} and ATCA observing the locations of 16 short GRB host galaxies and detected one. 

There are various possible reasons for, and contributions to, the 1.3~GHz flux measured for these GRB host galaxies candidates. If we assume that the observed emission can be attributed to star formation and not from an active galactic nucleus, it is typically dominated by two components: thermal bremsstrahlung around star-forming regions, and non-thermal synchrotron emission from cosmic-ray electrons accelerated in the galaxy's magnetic field. The latter is most dominant at low radio frequencies and can be related to the supernova rate in a galaxy, and through the supernova rate to the star formation rate. We can estimate an upper limit on the star formation rate in each host galaxy from the measured flux, using the formalism and equations from \citet{2011ApJ...737...67M}; and the results are shown in Table~\ref{tab:GRBhostSFR} as well. These star formation rates may be higher than star formation rates that can be derived from other observations in for instance the optical, because the radio observations provide a tracer that is unobscured by dust. Having said that, we consider the star formation rates in Table~\ref{tab:GRBhostSFR} to be upper limits, since there may be other contributions to the observed radio flux that are not taken into account in the star formation rate estimates we made, for instance the neutral hydrogen gas in these galaxies which is close to the centre of our observing band with MeerKAT. We can compare our measurements with \citet{2022ApJ...940...57N}, who used optical observations along with stellar populations to model star formation rates, as opposed to using radio observations. We can see that our rates are an order of magnitude higher on average. In particular, GRB 210323A and GRB 210919A have radio-derived star formation rates that are almost two orders of magnitude higher. This significantly higher rate could suggest that the emission from the neutral hydrogen gas is indeed a significant factor. However, the redshift of these GRBs pushes the emission from neutral hydrogen gas towards the lower end of the observing band, and in the case of GRB 210323A even outside of the observing band. This suggests that the star formation is in fact obscured in the optical.

\begin{table*}
\caption{GRB host galaxy locations, forced fit flux measurements (and one $3\sigma$ upper limit) from deep combined images, and star formation rates if the measured radio flux were completely attributed to star formation activity.}
\label{tab:GRBhostSFR}
\begin{tabular}{llllll}
\hline
        GRB & RA & Dec & Host Galaxy References & $F_{\rm{peak}}$ ($\mu$Jy)  & SFR ($M_{\odot}/yr$)\\
        \hline
        GRB200219A & 342.63795 & -59.11988 & \citet{2021AAS...23723503S,2022ApJ...940...56F} & $20\pm4$  &$12\pm2$\\
        GRB200411A & 47.66306 & -52.31654 & \citet{2021AAS...23723503S,2022ApJ...940...56F} & $46\pm3$ &$48\pm3$\\
        GRB200907B & 89.02896 & 6.90629 & \citet{2022ApJ...940...56F} & <22   &<18\\
        GRB210323A & 317.94717 & 25.36944 & \citet{2022ApJ...940...56F} & $18\pm6$  &$28\pm9$\\
        GRB210919A & 80.25814 & 1.31112 & \citet{2021AAS...23723503S,2022ApJ...940...56F} & $67\pm16$  &$11\pm2$\\
\end{tabular}
\end{table*}

\section{Discussion}
~\label{sec:discussion5}

\subsection{Comparison with Other Studies}

Comparing our upper limits on the afterglow emission with the upper limits in e.g., \cite{2015ApJ...815..102F} and \citet{2021ApJ...906..127F} shows that our limits are about as deep as the deepest ones in other short GRB studies; and in fact, our limits are close to many of the low-level detections of some short GRBs. Our observations are complementary to those other studies since we observe every GRB in the L-band, i.e., at 1.3~GHz, in contrast with the higher observing frequencies used in many other studies. Furthermore, since we carry out our multi-epoch observations to at least one month, no matter if the GRB was detected or not in earlier observations, we are able to narrow down the observational parameter space.

For most of the short GRBs with detections, the peak of the radio emission at 5.5 and 9\,GHz
 is within the first week \citep{2021ApJ...906..127F}. That peak should have been caught in 1.3~GHz observations out to about 1 month after the initial gamma-ray trigger, so our sample should not be missing such a light curve peak in our MeerKAT observations. Additionally, in the ATCA 5.5 and 9~GHz observations, all but one of the observations began within a day of the initial gamma-ray trigger. Furthermore, the \textit{e}-MERLIN 4.8~GHz observations started within two days of the trigger and for GRB 210323A continued almost daily out to about 10 days post-trigger. We note that we did not detect GRB~200522A at 1.3~GHz, which was detected at higher radio frequencies \citep{2021ApJ...906..127F}, but this may be due to a significantly higher noise level in our MeerKAT images due to nearby bright sources in the field. From the measurements given by \citet{2021ApJ...906..127F}, $\Psi$ is between 0.014 and 0.018. Comparing this with our $\Psi$ limits in Table~\ref{tab:psiobs1} reveals that our first observation has lower limits that are consistent with the detections, but the later observations give higher limits on $\Psi$. This implies that our first observation was not sensitive enough to detect the GRB afterglow, but the following ones should have detected it if the detections from \citet{2021ApJ...906..127F} were indeed emission from the forward shock. We also note that GRB~230217A was detected at higher frequencies as reported in~\citet{anderson23GCN}. From these measurements, along with a redshift of 0.533 taken from the average of the other GRB redshifts in this study, we calculate a $\Psi$ value ranging from 0.016 to 0.025. Comparing this to our lower limits in Table~\ref{tab:psiobs1} shows that they are much higher, again implying that the detection reported in~\citet{anderson23GCN} may not be produced in the forward shock. However, full broadband modeling would be needed to further investigate the discrepancy between the reported detections and our marginal or non-detections.

\subsection{Implications on GRB Parameters and Environments}

As described in Section~\ref{sec:methodandresult5}, our deep upper limits can be used to constrain some of the physical parameters that can be derived from peaks in radio light curves and spectral energy distributions. When comparing the $\Psi$ values and distributions from \citet{2017MNRAS.472.3161B} and \citet{2023MNRAS.518.1522D} with our lower limits on $\Psi$ using our ATCA, \textit{e}-MERLIN, and MeerKAT observations, as shown in Tables~\ref{tab:psiobs1}, \ref{tab:psiobs2}, and \ref{tab:psiobs3} along with Figure~\ref{fig:psiobshist}, it can be seen that our lower limits for the short GRBs are around the same values as the detections for a large long GRB sample. Also note in Table~\ref{tab:averagepsi} that the average in log-space of our $\Psi$ values are quite close to the values from \citet{2017MNRAS.472.3161B} and \citet{2023MNRAS.518.1522D}. Additionally, the lower limits on $\Psi$ calculated from only the ATCA observations have quite a few lower limits that are close to the upper end of the values shown in \citet{2017MNRAS.472.3161B} and \citet{2023MNRAS.518.1522D}. This means that either our noise levels in most of the images are very close to what is needed for having detections of these short GRBs, or the physical parameters in short GRBs are different than those for long GRBs. We will first discuss the latter option, and in the next subsection we will further discuss radio detectability of short GRBs.

The value of $\Psi$ strongly depends on two parameters related to electron acceleration: it has a linear dependence on $\epsilon_{\rm{e}}$, and is proportional to $\xi_{\rm{e}}^{-3/2}$. Both of these physical parameters are efficiency factors related to how relativistic shocks accelerate electrons. Since both long and short GRBs have relativistic shocks at the front of their jets, and one may expect the microphysics of this process to be the same or similar for all relativistic shocks, it would be unexpected for $\epsilon_{\rm{e}}$ and $\xi_{\rm{e}}$ to be very different between short and long GRBs. Therefore, we will explore other physical parameters that may be different for the two GRB classes.

While $\Psi$ depends strongly on $\epsilon_{\rm{e}}$ and $\xi_{\rm{e}}$, it depends weakly on other physical parameters. Most notably, $\Psi$ depends on the gamma-ray efficiency $\epsilon_{\gamma}$ and the circumburst density $n_0$ as $\Psi_{ISM}\propto n_0^{-1/4} ((1-\epsilon_{\gamma})/\epsilon_{\gamma})^{-1/4}$. 
Previous studies have presented conflicting conclusions on the typical values of the gamma-ray efficiency of GRBs and whether they differ significantly between long and short GRBs. Some studies \citep{10.1111/j.1365-2966.2006.10280.x,10.1093/mnras/stv2033,10.1093/mnras/stw1331} suggest that $\epsilon_{\gamma}$ has a fairly narrow distribution around $\sim0.15$. However, based on multi-wavelength modeling of a sample of long and short GRBs, \citet{2022MNRAS.511.2848A} find that the gamma-ray efficiency of short GRBs is significantly lower than for long GRBs, potentially by orders of magnitude. Deep radio observations could potentially distinguish between these two possibilities. If we look again at the parameter scalings for $\Psi$, we can see that $\Psi\propto (\epsilon_{\gamma}/n_0)^{1/4}$ for $\epsilon_{\gamma} << 1$. This means that for keeping $\Psi$ roughly the same, any change in $\epsilon_{\gamma}$ must have a roughly similar change in the density, or at least there cannot be a difference of orders of magnitude. In other words, if the gamma-ray efficiency is lower by orders of magnitude, as suggested by \citet{2022MNRAS.511.2848A}, the density will have to be lower by orders of magnitude as well, or otherwise the lower limits on $\Psi$ we find in our study are violated. In \citet{2022MNRAS.511.2848A}, however, the densities of the circumburst media are relatively high, similar to the densities for long GRBs.

Another point to consider is the scaling of the peak flux with the density, i.e., the peak flux being proportional to $n_0^{1/2}$. Using the aforementioned scaling relations, we can conclude that if the physical parameters in short and long GRBs are the same, including the densities and gamma-ray efficiencies, our study should be close to or marginally detecting these short GRBs since our lower limits for $\Psi$ (shown in Table~\ref{tab:psiobs1} and \ref{tab:psiobs2}) are approximately the same as the values of $\Psi$ in ~\citet{2023MNRAS.518.1522D}. If instead the gamma-ray efficiency of short bursts is much lower, this would require the densities to be lower as well, in which case we could still be close to detecting the short GRB afterglows. However, if the efficiencies are the same and the densities are much lower, the actual $\Psi$ values will be much higher than our lower limits (and therefore in agreement). 

Our sample of deep upper limits seems to show that there may be some differences in the physical parameters between long and short GRBs. Our potential detection of GRB 230217A at a relatively low flux value also provides some further hints about these parameters. It suggests that we should be close to detecting many of these sources, and that the sensitivity required to detect a much larger fraction of short GRB afterglows could be less than an order of magnitude. If this is true, it would suggest that perhaps there could be a difference in the gamma-ray efficiencies. Caution is warranted however, since this difference needs to be verified with further observations of future short GRBs. New and upgraded facilities may help resolve these different scenarios.

\subsection{Guiding Future Observations Towards Detections}

In the previous section, we discussed two scenarios in which we could be close to detecting short GRB afterglows in our observations. Our combinations of a potential detection along with non-detections suggests that gamma-ray efficiencies may indeed differ between long and short GRBs, as suggested by \citet{2022MNRAS.511.2848A}. However in order to verify this, next generation observatories, such as the Square Kilometer Array \citep[SKA;][]{2009IEEEP..97.1482D} or the Next Generation Very Large Array \citep[ngVLA;][]{2018ASPC..517....3M}, should be able to detect these sources. The first phase of the SKA, i.e., SKA-1, will have a sensitivity of about half an order of magnitude better in L band than MeerKAT, which may be sufficient for detecting these GRBs and allow for a more detailed analysis to determine all physical parameters from detailed radio analyses and multi-wavelength modeling. Additionally, ATCA will soon be upgraded with a new correlator that will provide wider bandwidth, and consequently lower noise for future observations. As discussed earlier, our ATCA observations provide excellent limits on $\Psi$, with many of our lower limits approaching the highest values of $\Psi$ show in \citet{2017MNRAS.472.3161B} and \citet{2023MNRAS.518.1522D}. These tight constraints on $\Psi$ imply that we should be very close to detecting these short GRB afterglows, thus with even modest improvements in sensitivity we expect to detect a much larger sample of these short GRB afterglows. 

We also discussed the possibility that our lower limits on $\Psi$ indicate much lower densities for short GRBs. In this case, it could be that the full SKA or the ngVLA would be required to start probing the full observational parameter space of short GRBs in the radio. In all cases, further observations with SKA-1 will disentangle these scenarios: if a thorough follow-up campaign of short GRBs with SKA-1 is performed and regularly detects them, we can really probe the physical parameter space; but if the result is mostly non-detections, we know that the gamma-ray efficiencies are the same between long and short bursts and that the densities must be very low, as indicated by some multi-wavelength modeling efforts \citep[e.g.,][]{2015ApJ...815..102F}. 

\section{Conclusion}
~\label{sec:conclusion5}

We have presented 38 observations of 8 short GRBs and measured deep upper limits on their radio afterglows at 1.3~GHz with MeerKAT, 11 observations at both 5.5 and 9~GHz for 4 of these short GRBs with ATCA, and 9 observations at 4.8~GHz of two GRBs with \textit{e}-MERLIN with 8 of these observations being of GRB~210323A. Due to the multi-epoch observations on a fairly dense, quasi-logarithmic cadence over the first month after the initial GRB trigger, we can constrain the peaks of the light curves at these radio frequencies, with in one case a potential detection. We use these limits to constrain the parameter $\Psi$, which provides insights in the microphysics of GRB afterglows and the densities of their environments. Using these constraints, and assuming that the physical parameters of long and short GRBs are similar, we estimate that we are very close to or marginally detecting these afterglows. We explored the possibility that some parameters may be different between short and long GRBs, for instance the gamma-ray efficiency or circumburst density. We show that a future survey of short GRBs with an upgraded ATCA and/or SKA-1 can help with determining whether or not the gamma-ray efficiency differs between long and short GRBs. If we continue to find mostly non-detections in a search with ATCA or SKA-1, we need to go even deeper with the full SKA or the ngVLA to probe the role of densities of the environments for short GRBs.

We also presented detections of possible host galaxies of 4 of the short GRBs in our sample, and constrained the star formation rate in those galaxies. Future studies comparing these measurements to observations at other wavelengths may be able to determine further properties of the host galaxies for these short GRBs.

\section*{Acknowledgements}

SIC would like to thank Brendan O'Connor for sharing optical images for the GRB fields and providing input on short GRB host galaxies. 
The authors would like to thank the operators, SARAO staff and ThunderKAT Large Survey Project team for all their efforts in supporting this project. 
The MeerKAT telescope is operated by the South African Radio Astronomy Observatory (SARAO), which is a facility of the National Research Foundation, an agency of the Department of Science and Innovation. 
We acknowledge use of the Inter-University Institute for Data Intensive Astronomy (IDIA) data intensive research cloud for data processing. IDIA is a South African university partnership involving the University of Cape Town, the University of Pretoria and the University of the Western Cape. We also acknowledge the computing resources provided on the High Performance Computing Cluster operated by Research Technology Services at the George Washington University. 

We acknowledge the Gomeroi people as the traditional owners of the ATCA observatory site. The ATCA is part of the Australia Telescope National Facility (\href{https://ror.org/05qajvd42}{https:// ror.org/ 05qajvd42}), which is funded by the Australian Government for operation as a National Facility managed by CSIRO. 
This paper includes archived data obtained through the Australia Telescope Online Archive ( \href{http://atoa.atnf.csiro.au}{http://atoa.atnf.csiro.au}).
GEA is the recipient of an Australian Research Council Discovery Early Career Researcher Award (project number DE180100346) funded by the Australian Government.

This work made use of data supplied by the UK \textit{Swift} Science Data Centre at the University of Leicester and the Swift satellite.
\textit{Swift}, launched in 2004 November, is a NASA mission in partnership with the Italian Space Agency and the UK Space Agency. \textit{Swift} is managed by NASA Goddard. Penn State University controls science and flight operations from the Mission Operations Center in University Park, Pennsylvania. Los Alamos National Laboratory provides gamma-ray imaging analysis.
This research made use of NASA’s Astrophysics Data System. This research has made use of SAOImage DS9, developed by the Smithsonian Astrophysical Observatory.
This research has made use of the VizieR catalogue access tool, CDS, Strasbourg, France \citep[DOI: 10.26093/cds/vizier;][]{vizier}.
A.H. is grateful for the support by the Israel Science Foundation (ISF grant 1679/23) and by the the United States-Israel Binational Science Foundation (BSF grant 2020203).
\section*{Data Availability}
Both image and calibrated visibility data is available upon request by email to sarahchastain1@unm.edu. Additionally, the data is available from the SARAO archive.



\bibliographystyle{mnras}
\bibliography{example} 





\bsp	
\label{lastpage}
\end{document}